\documentclass[twocolumn,preprintnumbers,amsmath,amssymb,nobibnotes,nofootinbib]{revtex4}

\usepackage{graphicx,epsfig}
\usepackage{subfigure}
\usepackage{psfrag}
\usepackage{dcolumn}
\usepackage{mathrsfs,amsmath}
\usepackage{mathrsfs}
\usepackage{slashed}
\usepackage[english]{babel}
\usepackage[usenames,dvipsnames]{color}
\RequirePackage[colorlinks=true
,urlcolor=blue
,anchorcolor=blue
,citecolor=blue
,filecolor=blue
,linkcolor=blue
,menucolor=blue
,pagecolor=blue
,linktocpage=true
,pdfproducer=medialab
]{hyperref}
\newcommand{\be}{\begin{equation}}
\newcommand{\ee}{\end{equation}}
\newcommand{\bea}{\begin{eqnarray}}
\newcommand{\eea}{\end{eqnarray}}

\newcommand{\ba}{\begin{array}}
\newcommand{\ea}{\end{array}}
\newcommand{\balg}{\begin{align}}
\newcommand{\ealg}{\end{align}}

\newcommand{\LL}{\mathscr{L}}
\newcommand{\cO}{\mathcal{O}}
\def\cy{{\bf y}}

\newcommand{\TeV}{\;\text{TeV}}
\newcommand{\GeV}{\;\text{GeV}}

\newcommand{\cms}{\;\text{cm}^3/\text{s}}

\newcommand{\hc}{\text{h.c.}}

\newcommand{\sv}{\sigma v}

\newcommand{\sdmp}{\sigma_{DM\, p} }

\begin{document} 

%
%
\preprint{FTUAM-13-129}
\preprint{IFT-UAM/CSIC-13-018}
\preprint{CERN-PH-TH/2013-034}

\title{Dark matter within the minimal flavour violation ansatz}

\author{Laura Lopez-Honorez}
\email{llopezho@vub.ac.be}
\affiliation{Theoretische Natuurkunde, Vrije Universiteit Brussel and The International Solvay Institutes Pleinlaan 2, B-1050 Brussels, Belgium}

\author{Luca Merlo}
\email{luca.merlo@uam.es}
\affiliation{Instituto de F\'{\i}sica Te\'orica UAM/CSIC and Departamento de F\'isica Te\'orica,\\
Universidad Aut\'onoma de Madrid, Cantoblanco, 28049 Madrid, Spain}
\affiliation{CERN, Department of Physics, Theory Division CH-1211 Geneva 23, Switzerland}
\date{\today}
\begin{abstract}
 Minimal Flavour Violation hypothesis can provide an attractive
 framework for Dark Matter (DM). We consider scalar DM candidates
 carrying flavour quantum numbers and whose representation under the
 flavour group guarantees DM stability. They interact with the
 Standard Model fields through Higgs portal at renormalisable level
 and also to quarks through dimension-6 operators.  We provide a
 systematic analysis of the viable parameter space for the DM fields, which are triplet
 of the flavour group, considering several DM-quark interactions. In this framework, we analyse in which cases
 the viable parameter space differs from Higgs portal models thanks to
 the underlying flavour structure. In contrast to minimal Higgs
 portal scenarios, we find that light DM in the GeV
 mass range as well as heavier candidates above Higgs resonance could
 be allowed by colliders, direct and indirect DM detection
 searches as well as flavour constraints. The large mass regime above $m_t$ could even be beyond the reach of
 future experiments such as Xenon 1T.
\end{abstract}

\maketitle

%
%
\section{Introduction}
\label{sec:introduction}
 
The determination of the nature of DM is one of the main challenges of 
cosmology and particle physics community today. The Standard Model
(SM) of particle physics, despite its great successes along with the
likely recent unveiling of its last building block at
LHC~\cite{:2012gk,:2012gu}, fails in providing a satisfactory DM
candidate.  In order to characterise the properties of the DM
particle, we are forced to go beyond the SM. As of today, no clear
scenario for New Physics (NP) is emerging. Concerning DM, we still
ignore its intrinsic properties, how it couples to the SM particles
and if the dark sector is made of one or several species.

Within the SM, the SM scalar or Brout-Englert-Higgs
(``Higgs'' for short) boson enjoys a special status since it can allow
for a direct coupling to the dark sector at renormalisable level. It
is already well known that the so-called ``Higgs portal'' provides a
quite simple and attractive framework for DM
phenomenology~\cite{Patt:2006fw,Kim:2006af,MarchRussell:2008yu,Kim:2008pp,Ahlers:2008qc,Feng:2008mu,Andreas:2008xy,Barger:2008jx,Kadastik:2009ca,Kanemura:2010sh,Piazza:2010ye,Arina:2010an,Low:2011kp,Djouadi:2011aa,Englert:2011yb,Kamenik:2012hn,Gonderinger:2012rd,Lebedev:2012zw}.
Interestingly, Higgs searches offer complementary
bounds to direct and indirect DM detection searches on the viable
parameter space of such DM models, especially in the low  mass
region~\cite{Farina:2011bh,Raidal:2011xk,Baek:2011aa,Djouadi:2011aa,Djouadi:2012zc,Andreas:2010dz,Gustafsson:2012aj,Mambrini:2012ue,Wang:2012ts}.

 The DM stability requirement is usually ensured by imposing by hand a
 discrete $Z_2$ symmetry, under which the DM candidate is odd while
 the SM fields are even. Several works have however been investigating
 which models of NP could provide a more fundamental origin for its
 stability, see e.g. Ref.~\cite{Hambye:2010zb} for a short
 review. A particular remark should be make on models where more sophisticated global discrete symmetries \cite{Ma:2008ym,Hirsch:2010ru,Haba:2010ag,Meloni:2010sk,Kajiyama:2010sb,Boucenna:2011tj,Adulpravitchai:2011ei,Meloni:2011cc,deAdelhartToorop:2011ad,Lavoura:2011ry,Hashimoto:2011tn,Boucenna:2012qb,Lavoura:2012cv,BenTov:2012tg,Bhattacharya:2013mpa} and horizontal gauge symmetries \cite{Hamaguchi:1998wm,Hamaguchi:1998nj,Batell:2010bp,Blennow:2010qp} have been adopted to deal with the DM phenomenology.
 
 Recently, Ref.~\cite{Batell:2011tc} sketched the typical
 features of DM candidates coupling to quarks in a way that
 is consistent with the Minimal Flavour Violation (MFV)
 ansatz~\cite{Chivukula:1987py,Hall:1990ac}.  Interestingly, the MFV,
 originally motivated to suppress dangerous flavour changing neutral
 current (FCNC) processes in NP contexts, can have a dual purpose
 guaranteeing the stability of the DM.
 
MFV is a working context that has been codified in
 Refs.~\cite{D'Ambrosio:2002ex,Cirigliano:2005ck,Davidson:2006bd,Gavela:2009cd,Feldmann:2009dc,Alonso:2011yg,Alonso:2011jd,Alonso:2012fy}
 as a general framework built upon the flavour symmetry of the kinetic
 terms of the SM Lagrangian. Focusing on the quark sector only, the
 latter presents a global flavour symmetry whose non-Abelian part is given by
\begin{equation}
G_f=SU(3)_{Q_L}\times SU(3)_{U_R}\times SU(3)_{D_R}\,.
\label{eq:Gf} 
\end{equation}
$G_f$ is only broken by the Yukawa interactions, unless the Yukawa
matrices are promoted to be auxiliary fields, called spurions,
transforming non-trivially under $G_f$. Without entering into
  details of a specific NP scenario, generic effects of flavour and CP
  violation can be described by means of an effective Lagrangian. If
  all the effective operators of this Lagrangian are constructed by
  means of the SM fields and the Yukawa spurions, all the flavour
  bounds are satisfied with a NP scale of few
  TeV~\cite{D'Ambrosio:2002ex,Cirigliano:2005ck,Davidson:2006bd,Fitzpatrick:2007sa,Kagan:2009bn,Gavela:2009cd,Lalak:2010bk,Grinstein:2010ve,Alonso:2011jd,Buras:2011wi,Alonso:2012jc,Alonso:2012px,Alonso:2012pz}.

A stable DM candidate can arise within the MFV context when the DM is
assumed to carry flavour quantum numbers (see also
Refs.~\cite{Arnold:2009ay,Agrawal:2011ze,Kamenik:2011nb}) and the DM
field representation under $G_f$ is chosen in order to prevent DM
decay into SM particles. Following the approach of
Ref.~\cite{Batell:2011tc}, we assume that the DM fields are neutral
under the SM gauge group, but transforms under $G_f$.  For simplicity,
we consider new scalar degrees of freedom as DM candidates with the
lowest representation under $G_f$ that guarantees their stability. By
construction, the DM flavour-multiplets do not couple at the
renormalisable level to quarks. Their interactions with SM particles
result from the interplay between Higgs portal interactions and
effective dimensions-six operators coupling DM to quarks. Notice that
Ref.~\cite{Batell:2011tc} proposed the rules for stable flavoured
DM. In that paper, the DM analysis focused on scalar DM interacting
through one particular $d=6$ operator and with a fixed hierarchy in
the DM components. The authors provided a survey of the typical bounds
that have to be taken into account from DM and flavour physics. This
study missed though several new features in the DM viable parameter
space that distinguish the DM within the MFV context compared to the
standard Higgs portal DM scenarios.

In this letter, our purpose is to perform a systematic analysis of the
viable parameter space of such DM context and to study how it differs
from the simple singlet scalar DM interacting via Higgs portal. In particular, we obtain new features in
the few GeV mass range,
thanks to DM coannihilation induced by flavour symmetry, as well as
larger viable parameter space for the DM mass range above the top mass
$m_{DM}>m_t$, when an extra interaction to $u-$type quarks is considered.

In the following, we first briefly summarise in Sec.~\ref{sec:mod}
the MFV context and its extension involving a potential DM candidate. In
Sec.~\ref{sec:EffOps}, we review the general form of effective $d=6$
operators providing DM-quark interactions compatible with the flavour
symmetry. In Sect.~\ref{sec:ParamSpace}, we study the parameter space
of this DM scenario considering bounds from direct and indirect DM
 and collider searches and  flavour constraints, and conclude in
Sec.~\ref{sec:Conc}.

\section{The MFV-DM context}
\label{sec:mod}

Deviations from the SM predictions induced by NP with generic flavour
structure are already severely constrained by the increasing accuracy
in the determination of the CKM matrix elements, the measurement of a
large number of FCNC processes and of CP asymmetries.  In a general
context, the scale of NP responsible for the flavour interactions
should be above hundreds or thousands TeV~\cite{Isidori:2010kg}.  On
the other hand, under the MFV hypothesis which  prescribes that all the sources of flavour and CP violation in any NP scenario are the
same as in the SM, the scale of NP responsible for the flavour
interactions is lowered down to few TeV.

This is technically implemented by constructing effective operators describing flavour and CP violation by
means of quarks and Yukawa spurions, that transform under the flavour
group $G_f$ in Eq.~(\ref{eq:Gf}). The $SU(2)_L$ doublet $Q_L$ and
singlets $u_R$ and $d_R$ transform according to 
\be
Q_L\sim({\bf3},{\bf1},{\bf1})\,,\quad
u_R\sim({\bf1},{\bf3},{\bf1})\,,\quad d_R\sim({\bf1},{\bf1},{\bf3})\,,
\end{equation} 
while the Yukawa spurions transform as
\begin{equation}
Y_u\sim({\bf3},{\bf\bar3},{\bf1})\,,\qquad
Y_d\sim({\bf3},{\bf1},{\bf\bar3})\,,
\label{YTransf}
\end{equation}
ensuring the invariance under $G_f$ of the Yukawa Lagrangian,
\be
\LL_Y=-\bar Q_L\,\tilde H\, Y_u\, u_R-\bar Q_L\,H\, Y_d\, d_R+\hc\,.
\ee
Quark masses and mixings are then correctly reproduced (but not predicted \cite{Alonso:2011yg}) once these
spurion fields get background values as 
\begin{equation} 
Y_u=\cy_u\,,\qquad
Y_d=V\,\cy_d\,, 
\end{equation} 
where $\cy_{u,d}$ are diagonal matrices whose elements are the Yukawa
eigenvalues, and $V$ is the CKM matrix.

Stable DM candidates can be found within the MFV framework looking for
the representations of $G_f$ that forbid the construction of operators
inducing the DM decay into SM degrees of freedom.  It has been
shown~\cite{Batell:2011tc} that the lowest representation under $G_f$
providing a stable DM candidate is a triplet under one
of the $SU(3)_i$ composing $G_f$. The DM stability is then insured for any
Lorentz representation of the DM candidate. For definiteness, we
consider scalar DM $S$, neutral under the SM gauge group, and focus on
the representation 
\begin{equation}
 S\sim({ \bf 3},{\bf 1 }, {\bf 1 })
\label{eq:Srepr}
\end{equation}
under $G_f$. Let us emphasise that our conclusions on the DM phenomenology
should apply to any scalar triplet under one $SU(3)_i$ term of $G_f$, after a slight modification of the dimension-six operators that drive its interactions to the quarks (see Sec.~\ref{sec:EffOps}). 
  
The low-energy effective Lagrangian describing our setup is given by
\begin{equation}
\LL = \LL_{SM} + \partial^\mu S^\dag \partial_\mu S - V(S, H)  + \LL^f_{\rm eff}+ \LL^{DM}_{\rm eff},
\label{FullLagrangian}
\end{equation}
where $\LL_{SM}$ is the SM Lagrangian, $V$ is the scalar
potential involving the DM field $S$ and the Higgs doublet $H$.
$\LL^f_{\rm eff}$ could contain $d=6$ pure-flavour operators,
described in e.g. Ref.~\cite{D'Ambrosio:2002ex}, that are suppressed
by $\Lambda_f^2$, while $\LL^{DM}_{\rm eff}$ contains $d=6$ DM-flavour
operators suppressed by $\Lambda^2_{DM}$.  $\Lambda_f$ and
$\Lambda_{DM}$ are the characteristic mass scales of the messengers of
the pure-flavour and DM-flavour interactions, respectively, and can a
priori be distinct. While $\Lambda_f$ should be larger than a few
TeV~\cite{D'Ambrosio:2002ex} to satisfy all flavour constraints, lower
values of $\Lambda_{DM}$ are still allowed. In practice, we do not explicitly
include the contribution of $ \LL^f_{\rm eff}$ for the numerical
analysis in Sec.~\ref{sec:numerical-analysis}, as $ \LL^f_{\rm eff}$ is not expected
to affect the DM phenomenology given that in our analysis $\Lambda_f$ is always larger than $\Lambda_{DM}$.

The SM scalar potential get modified at the renormalisable level by the
introduction of new scalar degrees of freedom. As already described in
Ref.~\cite{Batell:2011tc}, in the MFV context, the new contributions
to the scalar potential $V$ read 
\begin{equation}
  \begin{split}
V\supset &\, m_S^2 S^*_i (a\,\mathbf{1}_{ij}+ b\, (Y_u Y_u^\dag)_{ij}+\dots) S_j +\\ 
&+ \lambda \, S^*_i (a' \mathbf{1}_{ij}+ b' (Y_u Y_u^\dag)_{ij}+\dots)S_j\, H^\dag H , 
\end{split}
\label{portal}
\end{equation}
where $i,j$ are flavour indices, $a,b,a',b'$ are dimensionless
$\cO(1)$ parameters and the ellipsis denotes further
negligible\footnote{Here we follow the common practice considering
  all interactions up to the first power in $Y_uY_u^\dag$. See
  e.g. Ref.~\cite{Kagan:2009bn} for an analysis studying the impact of
  resuming over the Yukawa coupling expansion in the context of MFV
  in the absence of DM. }  $Y_d$ spurion insertions.  After flavour and electroweak symmetry breaking (EWSB), $\langle H \rangle = {\rm v}/\sqrt{2}$ with ${\rm v}=246$ GeV, the $S$ mass-squared matrix is
diagonal and given by \begin{equation} \LL_m\supset - S^*_i \left[
    m_A^2 + m_B^2\, \cy^2_{u_i} \right]S_i\,,
\label{spectrum}
\end{equation}
where we have defined 
\begin{equation}
m_A^2 =  m_S^2 a + \frac12 \lambda {\rm v}^2 a'\,,\qquad
m_B^2 =  m_S^2 b+ \frac12 \lambda {\rm v}^2 b'\,.
\end{equation}
Of phenomenological importance are the trilinear couplings of the DM particles to the physical Higgs $h$, which in the mass eigenbasis read:
\begin{equation}
\LL \supset - \frac12 \lambda\,  {\rm v}\, h\, {S}^*_i \left( a' + b' \cy^2_{u_i}\right) S_i \equiv - \frac12  \lambda_i\,  {\rm v}\, h\, {S}^*_i S_i,
\label{trilinear}
\end{equation}
where $\lambda_i \equiv\lambda(a'+ b' \cy^2_{u_i})$.

\section{The DM-Flavour Operators}
\label{sec:EffOps}

$S$ transforming as a triplet of $G_f$ as in Eq.~(\ref{eq:Srepr})
prevents the construction of any operator containing a single DM field
and quarks. A coupling between pairs of DM multiplets and quarks is
still allowed through non-renormalisable interactions suppressed by
$\Lambda_{DM}^2$.  When the flavour symmetry $G_f$ and
the EW symmetry are unbroken, one can consider the following $d=6$
operators:
\begin{equation}
\LL^{DM}_{eff}  =  \frac{1}{\Lambda_{DM}^2}\sum_{\alpha = 1}^{5} c^{\alpha}_{ijk\ell} \left({\cal O}_\alpha\right)_{ijk \ell}\,,
\label{eq:LDM}
\end{equation}
with
\begin{align}
\left({\cal O}_{1}\right)_{ijk\ell} & = (\bar Q_{L_i} \gamma^\mu Q_{L_j} )(S_k^*  \overleftrightarrow {\partial_\mu} S_\ell)\,, 
\label{op1}
 \\
\left({\cal O}_{2}\right)_{ijk\ell} & = (\bar u_{R_i} \gamma^\mu u_{R_j} )(S_k^*  \overleftrightarrow {\partial_\mu} S_\ell)\,, 
\label{op2}
 \\
\left({\cal O}_{3}\right)_{ijk\ell} & = (\bar d_{R_i} \gamma^\mu d_{R_j} )(S_k^*  \overleftrightarrow {\partial_\mu} S_\ell)\,,  
\label{op3}
\\
\left({\cal O}_{4}\right)_{ijk\ell} & = (\bar Q_{L_i} u_{R_j} )(S_k^*  S_\ell) \tilde H  +{\rm h.c.}\,, 
\label{op4}
\\
\left({\cal O}_{5}\right)_{ijk\ell} & = (\bar Q_{L_i} d_{R_j} )(S_k^*  S_\ell) H + {\rm h.c.}\,,
\label{op5}
\end{align}
where $i,j,k,\ell$ are flavour indices (see also Ref.~\cite{DelNobile:2011uf,Kamenik:2011vy}).  The
operators above correspond to an effective theory description of the
NP sector giving rise to an additional sources of quark-DM
coupling besides the Higgs portal interactions. The coefficients
$c^{\alpha}_{ijk\ell}$ take into account  all possible flavour
contractions and read as
\begin{align}
\begin{split}
c^1_{ijk\ell} =& c^1_1 \mathbf{1}_{ij}  \mathbf{1}_{k\ell} 
+ c^1_2 \mathbf{1}_{i\ell} \mathbf{1}_{kj} 
+ c^1_3 (Y_u Y_u^\dag)_{ij} \mathbf{1}_{k\ell}+ \\ 
&+ c^1_4  \mathbf{1}_{ij}  (Y_u Y_u^\dag)_{k\ell}
+ c^1_5 (Y_u Y_u^\dag)_{i\ell} \mathbf{1}_{kj} 
+ \dots
\end{split}
\label{c1}\\
\begin{split}
c^2_{ijk\ell}  = & c^2_1 \mathbf{1}_{ij}  \mathbf{1}_{k\ell} 
+ c^2_2 (Y_u^\dag Y_u)_{ij} \mathbf{1}_{k\ell} +\\
&+c^2_3  \mathbf{1}_{ij}  (Y_u Y_u^\dag)_{k\ell}+ c^2_4 (Y_u^\dag)_{i\ell}  (Y_u)_{kj}+ \dots
\end{split}
\label{c2}\\
\begin{split}
c^3_{ijk\ell}  = & c^3_1 \mathbf{1}_{ij}  \mathbf{1}_{k\ell}
+ c^3_2 (Y_d^\dag Y_d)_{ij} \mathbf{1}_{k\ell} + \\ 
&+c^3_3  \mathbf{1}_{ij}  (Y_u Y_u^\dag)_{k\ell}+ c^3_4  (Y_d^\dag)_{i\ell} (Y_d)_{kj}+ \dots
\end{split}
\label{c3}\\
\begin{split}
c^4_{ijk\ell} = & c^4_1 (Y_u)_{ij}  \mathbf{1}_{k\ell} 
+ c^4_2 \mathbf{1}_{i\ell} (Y_u)_{kj}+\\ 
&+c^4_3 (Y_u)_{ij}  (Y_u Y_u^\dag)_{k\ell}
+ c^4_4 (Y_u Y_u^\dag)_{i\ell} (Y_u)_{kj}+ \dots
\end{split}
\label{c4}\\
\begin{split}
c^5_{ijk\ell}  =& c^5_1 (Y_d)_{ij}  \mathbf{1}_{k\ell} 
+ c^5_2 \mathbf{1}_{i\ell} (Y_d)_{kj}+\\ 
&+c^5_3  (Y_d)_{ij}  (Y_u Y_u^\dag)_{k\ell}
+ c^5_4 (Y_u Y_u^\dag)_{i\ell} (Y_d)_{kj} + \dots
\label{c5}
\end{split}
\end{align}
where we have considered all possible terms up to first
powers in $Y_uY_u^\dag$ and the dots refer to
negligible contractions associated to $Y_d$ insertions. In the
following, we use real coefficients $c^{\alpha}_a$, according to the MFV ansatz
under which all the sources of flavour and CP violation should be
associated to the Yukawas only. 

The scale $\Lambda_{DM}$ in Eq.~(\ref{eq:LDM}) corresponds to a
function of couplings and mediator masses whose exact combination
depends on the UV completion of the effective theory under study. In
principle, different operators or flavour contractions can depend on
different type of messengers. $\cO_{1,2,3}$ would for instance
be the result of a vector boson exchange, while $\cO_{4,5}$ would
be due to the exchange of scalar or fermionic mediators.  Here we
assume that the structure of Eqs.~(\ref{eq:LDM})-(\ref{c5}) captures the DM phenomenology as long as the DM mass is lighter
than the particles mediating the interactions. 

Also notice that when the energy scale involved in the physical
processes is below the EWSB scale, both the gauge $SU(2)_L\times U(1)_Y$ and the flavour $G_f$ symmetry descriptions used in the definition of Eqs.~(\ref{op1})-(\ref{op5}) break down. One could then wonder for the appearance of new kind of operators, especially for those inducing DM decay, but this is not the case: the effective
theory at low energy must match the theory at higher energies and therefore no new gauge or flavour couplings appear.

\section{The viable parameter space}
\label{sec:ParamSpace}

In this section, we study the viable parameter space for DM analysing
the impact of each dimension-6 operator $\cO_{1,..,5}$ on DM
phenomenology.
\subsection{Constraints from DM and flavour physics}
\label{sec:constraint}
  Let us first summarise the constraints that have been imposed in
  each scenarios. We consider that $\lambda_i<\pi $ in
  order to preserve a perturbative regime. We take
  $\lambda_i>-\sqrt{\pi \lambda_h} $, where $ \lambda_h$ is the Higgs
  self coupling, to ensure that the scalar potential is bounded from
  below, assuming that the DM scalar self-couplings are up to
  $\sim~\cO(\pi)$. We also impose that $\Lambda_{DM}>m_{S_1},
  m_{S_2},m_{S_3}$ to avoid breaking down of the effective field
  theory description. In addition, from DM and flavour physics, we
  have to take into account the following points:
 \begin{enumerate}
 \item \label{item:relic}{\bf \it Relic abundance $\Omega_{DM}$:} The
   total DM abundance deduced from cosmological
   observations constrains the thermally averaged
   effective annihilation cross-section $\langle \sv \rangle$.  In the simplest
   cases~\cite{Griest:1990kh}, for the Taylor expanded annihilation cross section times the center of mass velocity
   $\sigma v=a+b\,v^2$, the Boltzmann equations for the evolution of
   DM number density gives rise to the relation
   \begin{equation}
     \Omega_{DM} h^2= \frac{1.7\times 10^9\,x_f  \,{\rm GeV^{-1}}}{\sqrt{g_*} (a+3\,b/x_f) m_{pl}}
     \label{eq:omdm}
   \end{equation}
  where $h$ is the Hubble parameter\footnote{Here, we keep the notation $h$
    for the Hubble parameter as it is widely used in the
    literature. It should not be confused with the physical Higgs field.},
  $x_f$ is the ratio between the DM mass and its freeze out
  temperature, $g_*$ is the number of relativistic degrees of freedom and
  $m_{pl}=1.22\times 10^{19}$ GeV is the Planck mass. The constraint
  $\Omega_{DM} h^2 \sim 0.1$~\cite{Hinshaw:2012fq} translates then
  into $\langle \sv \rangle \sim 3\times 10^{-26}$ cm$^3/s$ or
  equivalently $\langle \sv \rangle \sim 3\times 10^{-9}$
  GeV$^{-2}$. In our numerical analysis of
  Sec~\ref{sec:analytical-insights}, we use the code {\tt MicrOMEGAs}~\cite{Belanger:2008sj,Belanger:2010gh} that integrates more accurately the set of Boltzmann equations and we
  impose $ 0.09<\Omega_{DM} h^2<0.13$.

\item \label{item:ID-DD}{\bf \it Direct and indirect detection:}
  Direct detection searches are among the best test of Weakly
  Interacting Massive Particle (WIMP) DM scenarios. This is especially
  the case of spin independent DM-proton scattering and for masses
  around 50 GeV.  The associated cross-section $\sdmp$ should be below
  the bounds of experiments such as PICASSO~\cite{Archambault:2012pm}
  and Xenon 100~\cite{Aprile:2012nq} for $m_{DM}$ in the GeV-TeV
  range. Notice that such experiments have an energy threshold in the
  few GeV range and PICASSO's threshold is among the lowest
  ones.\\ 
  Indirect detection searches are also digging into the viable
  WIMP mass range.  The annihilation cross-section times center of
  mass velocity $\sv$, can be tested by FERMI
  experiment~\cite{Ackermann:2011wa}. The Cosmic Microwave Background
  (CMB) observation experiments such as WMAP~\cite{Hinshaw:2012fq}, SPT~\cite{Hou:2012xq} and
  Planck~\cite{Planck:2006aa} (in the very near future) provide complementary
  constraints on $\sv$. Indeed, the energy released into the Inter
  Galactic Medium (IGM) by DM annihilation can alter the thermal
  history of the Universe, leading to observable changes in CMB
  observables, see~\cite{Padmanabhan:2005es,Natarajan:2009bm,Cirelli:2009bb,Finkbeiner:2011dx}
  and \cite{Giesen:2012rp, Cline:2013fm} for the latest analysis.

\item \label{item:LHC}{\bf \it Colliders:} Assuming that the particle
  resonance with a mass of about 125 GeV observed by the ATLAS and CMS
  collaborations at the LHC~\cite{:2012gk,:2012gu} corresponds to the
  Higgs boson, one can constrain the invisible branching ratio to be
  $\mbox{Br}(h\to \mbox{ inv})<0.15$ at 2$\sigma$
  level~\cite{Giardino:2012dp}. Latest constraints from monojet events
  observed by the CMS~\cite{Chatrchyan:2012me} and
  ATLAS~\cite{ATLAS:2012ky} collaborations can give rise to stringent
  bounds on $\sdmp$ complementary to direct detection searches for
  WIMP masses in the GeV range~\cite{Beltran:2008xg,
    Beltran:2010ww}. Let us emphasise that the limits were derived for
  fermionic DM and in the case of scalar DM these bounds are not
  always that constraining. In the following, we use the results
  presented in Ref.~\cite{Goodman:2010ku}.

\item {\it Meson decays:} The bounds on meson decays into invisible
  final state (see e.g. Ref.~\cite{Kamenik:2011vy} for a review)
  strongly limit the direct couplings of quarks, apart from the
  top\footnote{ In Ref.~\cite{Kamenik:2011nb} it was argued that
    single top production at LHC with large missing transverse energy
    involving flavour violating interactions can give rise to novel
    signature for DM detection. Notice though that for the DM models
    studied here, the $t$ quarks only decay to $u$ and $c$ quarks and
    up to now LHC experiments can still not distinguish among light
    quarks jets.}, to the DM. In order to pass these constraints, one
  has to impose that $\Lambda_{DM}$ is larger than hundreds of GeV for
  low mass DM. Dimension-6 operators with such a large scale of NP can
  not guarantee the right DM relic abundance. Combining such bounds
  with the necessary small couplings $\lambda$ to avoid large Higgs
  invisible decay width, a DM particle coupling mainly to $u-$type and
  $d-$type quarks through $\cO_{1,..,5}$ with mass $m_{DM}<m_{D}/2$
  and $m_{DM}<m_{B}/2$, respectively, is then excluded.

\item {\it Meson oscillations:} Since no tree-level diagrams mediated
  by DM particles can be drawn, contributions to meson oscillations
  appear only at the loop level, as in the SM. The operators
  $\cO_{1,2,3}$ give negligible contributions to the meson oscillation
  observables: the MFV ansatz ensures that the GIM mechanism also
  holds for these loop diagrams; an additional suppression $\propto
  1/\Lambda_{DM}^4$ makes NP contributions smaller than the corresponding
  SM ones. In contrast, the operators $\cO_4$ and $\cO_5$
  have a LR chiral structure and the corresponding contributions to
  the meson oscillations are chiral enhanced. Following
  Ref.~\cite{Batell:2011tc}, the Wilson coefficients of the effective
  interactions $(\bar q_{R_i}\,q_{L_j})(\bar q_{R_i}\,q_{L_j})$, for
  $i>j$, depend on
  \begin{equation}
     C_{RL}^{ij}\propto\dfrac{m^2_{q_i}}{32\pi^2\Lambda_{DM}^4}\left(V^*_{tj}V_{tj}\right)^2F\left(\dfrac{m^2_{S_h}}{m^2_{S_l}}\right)\,,
²\label{eq:CRLij}
  \end{equation}
  where $m_{S_h} (m_{S_l})$ is the heaviest (lightest) of the $S$
  components and $F(x)=((x+1)\,\log x)/(x-1) - 2$ is the Inami-Lin function that
  vanishes in the limit $x\to1$. For $\Lambda_{DM}\geq {\rm v}$, these NP
  contributions can be neglected while for $\Lambda_{DM}\ll {\rm v}$, they
  can be large. Notice that for $K$ meson system, the Wilson
  coefficients are suppressed by $m^2_s/\Lambda_{DM}^2$ and no sizable
  effects can be seen in $\Delta M_K$ and $\epsilon_K$ (see Ref.~\cite{Isidori:2013ez} for a recent review for details on the meson oscillation observables). Let us also
  mention that for $B_d$ and $B_s$ systems, we expect new
  contributions in $\Delta M_{B_d}$ and $\Delta M_{B_s}$ only. Indeed,
  the absence of NP phases, in agreement with the MFV ansatz, prevents
  modifications on $S_{\psi K_S}$ and $S_{\psi\phi}$. In addition, let us
  stress that when the DM components are very near in mass the Wilson
  coefficient get an extra suppression through the Inami-Lin
  function. This is typically the case when the DM relic abundance is
  driven by coannihilations.

For the models considered in this paper, only the DM-quark
interactions through $\cO_{4,5}$ can thus contribute substantially to
meson oscillations. In principle, the operator $\cO_{4}$ can give rise
to non negligible contributions in $\bar D^0-D^0$ oscillations. The
large theoretical uncertainties in the $D$ meson system prevents
though to set relevant bounds on $\cO_{4}$ mediated interactions. This
is not the case in $B$ meson systems. In the presence of $\cO_5$
interactions, we impose that NP contributions from~(\ref{eq:CRLij})
are at most equal to the theoretical uncertainties on $\Delta M_{B_d}$
and $\Delta M_{B_s}$, i.e. $\Delta
C^{ij}=C_{RL}^{ij}/C_{SM}^{ij}\leq0.1$.

Let us also mention that, the operators $\cO_4$ and $\cO_5$ induce also modifications of the
  quark mass terms, when contracting the two DM legs at loop to draw a tadpole diagram: these
  contributions can be safely neglected, providing relative corrections suppressed by loop
  factors and by $m^2_{DM}/\Lambda^2_{DM}$. 
  \end{enumerate}

\subsection{Analytical insights}
\label{sec:analytical-insights}

As previously discussed, the Lorentz structure of the operators
$\cO_{1,..,5}$ can be categorised into two subgroups. The operators
$\cO_{1,2,3}$ could arise from vector boson exchange, while operators
$\cO_{4,5}$ would be associated to fermion or scalar exchange.  At
this point, it is instructive to analyse the scale of NP
$\Lambda_{DM}$ that would a priori be necessary in order to give rise
to the right relic abundance assuming a negligible contribution from
Higgs portal. This is particularly relevant for DM masses below the
Higgs resonance, where the bounds on the decay width of Higgs into
invisible final state strongly constrains the DM-Higgs couplings
$\lambda_i$ (see the colliders section in Sec.~\ref{sec:constraint}).

In the limit of low center of mass velocity, $v\to 0$, the annihilation cross-sections associated to the processes
$S_i^*S_i\to \bar q_jq_j$ for a fixed value of $i$ and $j$ become at leading order in $v$: 

\begin{eqnarray}
  \left.\sv\right|_{\cO_{1,2,3}} &\simeq&\frac{N_c\, c^2 v^2}{48\pi\Lambda_{DM}^4}\frac{m_q^4-5 m_q^2m_{DM}^2+4m_{DM}^4 }{m_{DM}^2\left(1-{m_q^2}/{m_{DM}^2}\right)^{1/2}} \cr
  \left.\sv\right|_{\cO_{4,5}}  &\simeq& \frac{N_c\, c'^2 m_q^2}{4\pi\Lambda_{DM}^4} \left(1-{m_q^2}/{m_{DM}^2}\right)^{3/2}
\label{eq:sv}
\end{eqnarray}
where $N_c$ is the number of colors of the final state quark, $c,c'$
are a combination of $\cO (1)$ coefficients and $m_{DM}$ and $m_q$ are
the masses of $S_i$ and $q_j$ respectively. The first cross-section is
p-wave suppressed while the second one corresponds to a s-wave driven
process. Eq.~(\ref{eq:omdm}) implies that for e.g. $m_{DM}=50$ GeV,
$\Lambda_{DM}$ should be $\sim 450$ GeV for $\cO_{1}$ (five families
of quarks are involved) and $\sim 200$ GeV for $\cO_{5}$ (mainly $b$
quarks involved) in order to get the right relic abundance through
annihilation only driven by the dimension-6 operators.  Let us
emphasise though that given the velocity dependence of the $\sv$'s
above, it is clear that prospects for indirect detection, involving
velocities $v\sim 10^{-3}$, will a priori be more constraining
for $\cO_{4,5}$ than $\cO_{1,2,3}$. In practice, the coannihilations
and Higgs portal interactions complicate the relic density
analysis. The latter processes are fully taken into account in our
numerical treatment of the DM models with the {\tt MicrOMEGAs } code
after having introduced the proper Feynman rules using the {\tt
  LanHEP} package~\cite{Semenov:2008jy}. We have also used {\tt
  MicrOMEGAs}'s tools for the calculation of the cross-sections
relevant for direct and indirect detection searches.

\subsection{Numerical analysis}
\label{sec:numerical-analysis}
We have performed a random scan of the viable parameter space for DM
within the following parameter ranges:
\begin{equation}
\begin{gathered}
0.3<|a,a',b,b',c^\alpha_a|<1.3\\
0.1 \GeV<m_{S_3}< 1 \TeV\\
m_{DM}<\Lambda_{DM}< 10 \TeV\\
10^{-5} <|\lambda|< \pi\,
\end{gathered}
\label{eq:ranges}
\end{equation}
considering real, positive and negative values for the parameters
$a,a',b,b',c^\alpha_a$ and $\lambda$. Let us emphasise that once the
value of $m_{S_3}$ is fixed, one can obtain larger or smaller values
of $m_{S_1}$ and $m_{S_2}$ by scanning over $a,a',b,b'$ and $\lambda$
parameters. As a result, in contrast to Ref.~\cite{Batell:2011tc}, we
have not imposed in our analysis that $S_3$ is the lightest component
of the $S$ triplet.  The constraints from perturbativity, stability of
the scalar potential, validity of the effective theory, and DM relic
abundance are applied during the scan (see Sec.~\ref{sec:constraint} for more details).

\begin{figure*}[!]
\hspace*{-0.4cm}
\subfigure{\includegraphics[width=6cm]{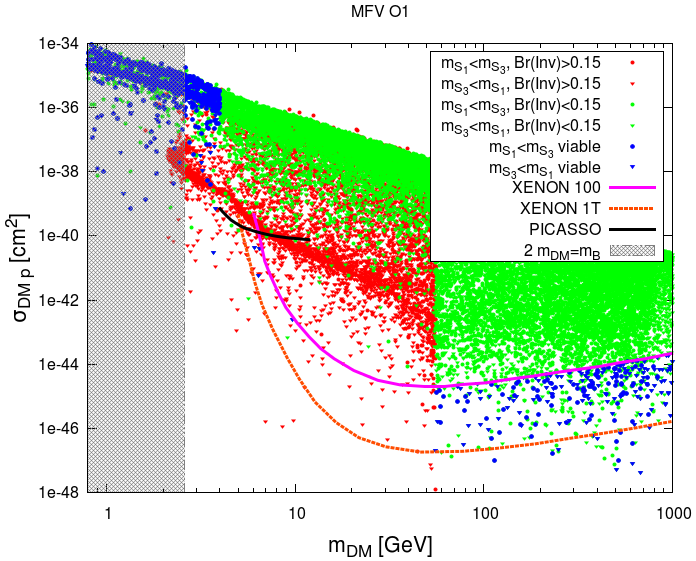}}
\subfigure{\includegraphics[width=6cm]{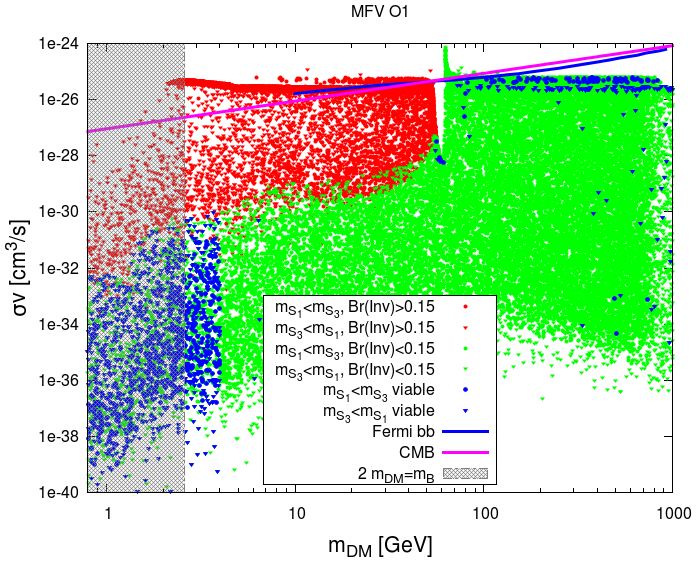}}
\subfigure{\includegraphics[width=6cm]{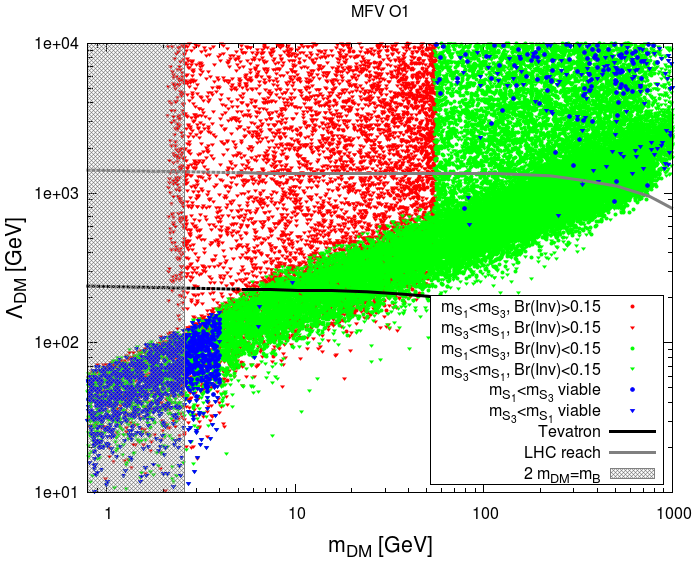}}
 \caption{Operator ${\cal O}_1$. Left: DM proton scattering
   cross-section as a function of the DM mass. Center: Annihilation
   cross-section as a function of the DM mass. Right: Scale of NP as a
   function of the DM mass. In all figures, the red points are
   excluded by the constraint on the Higgs invisible decay
   branching ratio, while blue (green) points (do not) pass the direct
   and indirect detection searches bounds.}
 \label{fig:O123}
\end{figure*}
\begin{figure*}[ht!]
\begin{center}
\subfigure{\includegraphics[width=8cm]{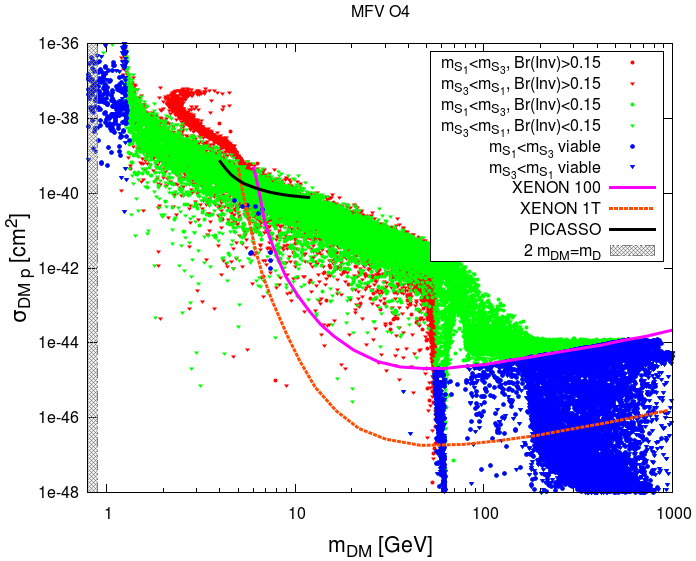}}
\subfigure{\includegraphics[width=8cm]{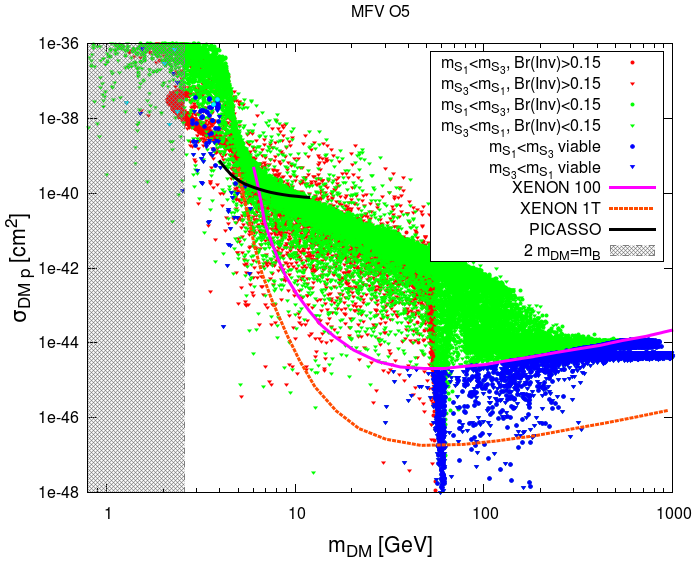}}\\
\subfigure{\includegraphics[width=8cm]{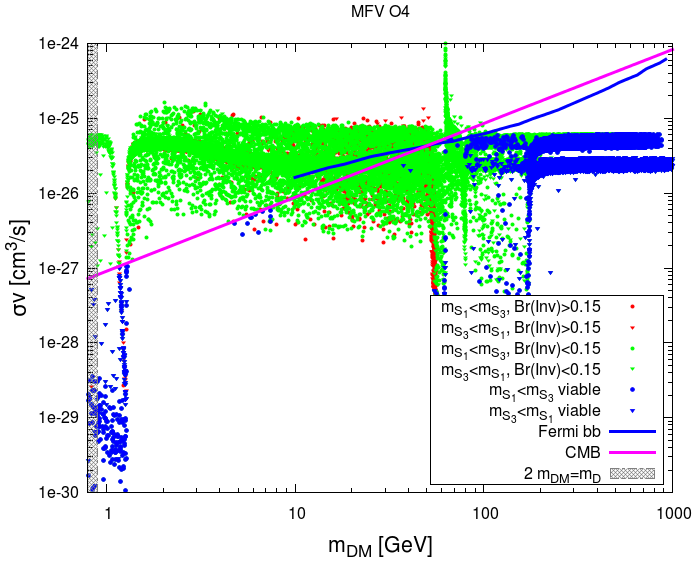}}
\subfigure{\includegraphics[width=8cm]{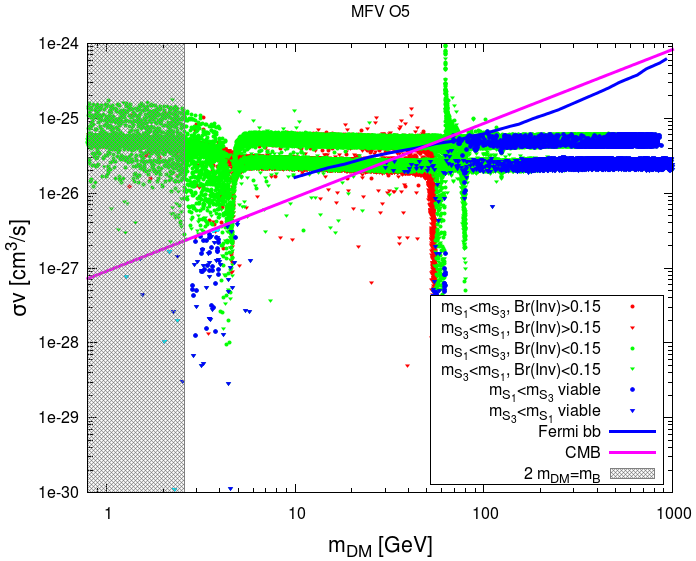}}\\
\subfigure{\includegraphics[width=8cm]{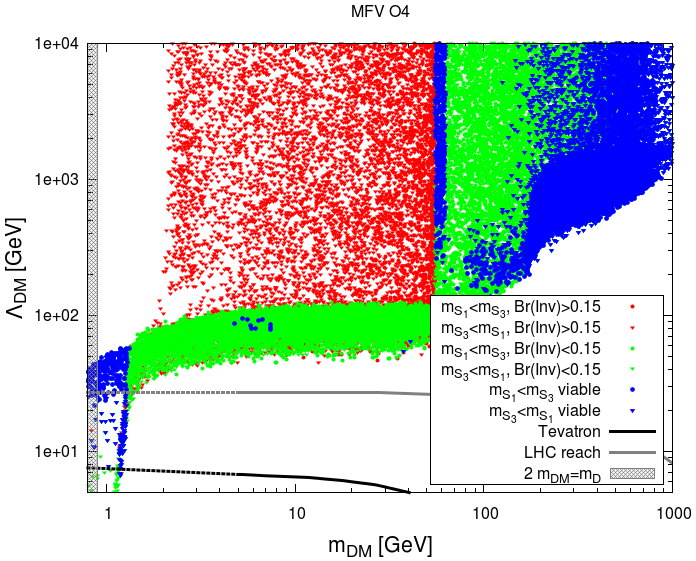}} 
\subfigure{\includegraphics[width=8cm]{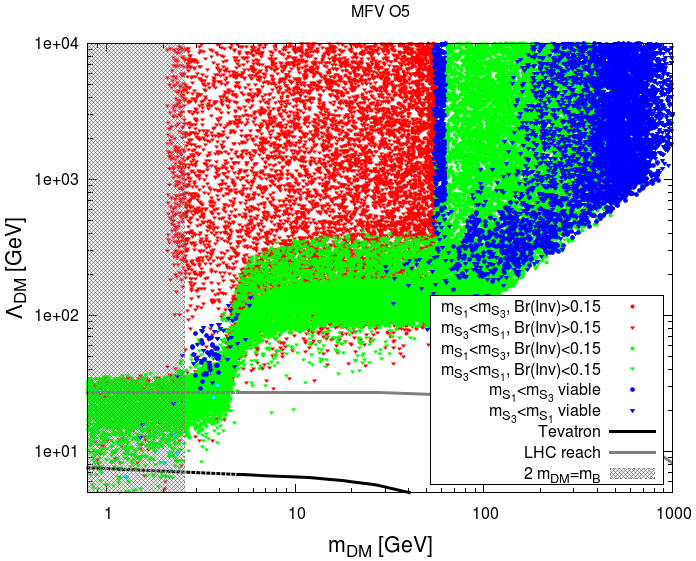}}
\caption{\label{fig:O45-s-m} Operators ${\cal O}_4$ (left) and ${\cal
    O}_5$ (right). Top: DM proton scattering cross-section as a
  function of the DM mass. Center: Annihilation cross-section as a
  function of the DM mass. Bottom:Scale of new physics as a function
  of the DM mass. In all figures, the red points are excluded by the
  constraint on the Higgs decay invisible branching ratio, while
  blue (green) points (do not) pass the direct and indirect detection
  searches bounds.}
\end{center}
\end{figure*}

In Figs.~\ref{fig:O123} and~\ref{fig:O45-s-m}, we present the results
of the scans for the operators $\cO_{1}$ and $\cO_{4,5}$ in the
planes $m_{DM}-\sdmp$, $m_{DM}-\sv$ and $m_{DM}-\Lambda_{DM}$
superimposing the constraints from direct, indirect and colliders
searches respectively. The corresponding plots for operators
$\cO_{2,3}$ are very similar to the ones for $\cO_{1}$. The red points
are excluded by constraints on the invisible decay branching ratio of
$125$ GeV Higgs, while the blue (green) points (do not) pass the
constraints from the combined PICASSO, Xenon 100, FERMI and CMB
experiments.

Notice that in the $m_{DM}-\sv$ plots, we only show the constraint
from FERMI on annihilation into $\bar b b$ at 100~\% as they are the
most limiting in the model under study. The limit associated to CMB
represented in the plots results from the analysis of
Ref.~\cite{Giesen:2012rp}. The latter makes use of the recent SPT data
assuming a value of the fraction of released energy in the IGM of 0.2
which should be representative of DM annihilating to
quarks~\footnote{Notice that a similar analysis~\cite{Cline:2013fm}
  shows the expected limits for each annihilation channels for WMAP7
  and future Planck data release. It appears that the limits do not
  vary much going from annihilations into quarks to annihilation into gauge or Higgs boson. Also notice
  that Ref.~\cite{Cline:2013fm} obtain a slightly less constraining
  bound for future Planck data. Here we stay a bit more conservative using the results of
  Ref.~\cite{Giesen:2012rp}.}.  In the legend of the plots, the
``viable'' reference should be interpreted as ``passes direct,
indirect detection and CMB constraints''. In the $m_{DM}-\sdmp$, we also
present the reach of the future Xenon 1T experiment~\cite{Xenon1T}
that will test most of the large mass regime of $\cO_{1,..,5}$ operators.

  We will see that some blue points do not pass monojet searches
  constraints from Tevatron or are within the reach of LHC at
  $\sqrt{s}=14$ TeV. In Figs.~\ref{fig:O123} and \ref{fig:O45-s-m}, we
  represent with dotted lines the continuation of the limits from
  Ref.~\cite{Goodman:2010ku} to masses $m_{DM}<$ 5 GeV. Notice however
  that in Ref.~\cite{Goodman:2010ku} slightly different assumptions
  were made on the coupling of DM to quarks compared to the present
  analysis. The lines in the $m_{DM}-\Lambda_{DM}$ associated to
  Tevatron and LHC constraints are thus mainly a guide for the eye.
  We thrust though that $\cO_{1,2,3}$ low mass candidates are already
  excluded by such analysis (see Fig~\ref{fig:O123}), while for
  $\cO_{4,5}$ only a few models are within the reach of LHC. A
  more detailed analysis of LHC prospects for detection of such a DM
  candidate is however beyond the scope of the present paper.

\subsubsection{ General features}
\label{sec:general-features}

Several features appear in Figs.~\ref{fig:O123} and \ref{fig:O45-s-m}.
Let us first concentrate on the $\sv-m_{DM} $ plots. Typically, when
the correct relic abundance is achieved and it is driven by
annihilation processes, the annihilation cross-section today reaches
$\sim 3\times 10^{-26}\cms$. In this $\sv$ range, we clearly see two
preferred values of the annihilation cross-sections, especially in
Fig.~\ref{fig:O45-s-m}. The annihilation cross-section for $S_{1,2}$
DM today appears to be typically twice as large as the one
for $S_3$.

 This can be understood considering the fact that, in the MFV context,
 $S_1$ and $S_2$ are generally quite degenerate, see
 Sec.~\ref{sec:mod}. In the latter case, the effective DM annihilation
 cross-section~\cite{Griest:1990kh} driving the relic abundance goes
 as $\langle \sv \rangle \sim (\sv_{11} +\sv_{12}+\sv_{21}
 +\sv_{22})/4$ where $\sv_{ij}$ refers to the cross-sections times
 velocity associated to $S_iS_j^*\rightarrow X$ annihilation and
 coannihilation processes ($X$ is a SM final state) in the $v
 \rightarrow 0$ limit, assuming that the relic abundance is driven by
 s-wave dominated processes. When coannihilation processes are
 suppressed, i.e. $\sv_{ij \, i\neq j}\ll \sv_{ii}$ and $\sv_{ii}$ is
 identical for $i=1,2$, we get $\langle \sv \rangle =\sv_{11}/2 \sim
 3\times 10^{-26}$ cm$^3$/s. This is what we observe in e.g. the upper
 plot of Fig.~\ref{fig:O45-s-m} when $\Lambda_{DM}$ is large and more
 especially when the interactions driven by $c_2^\alpha$ terms in
 Eqs.~(\ref{c4})-(\ref{c5}) are suppressed. Notice that we
 apply the detection searches constraints as if all the DM in the
 galaxies was made of only one species $S_i$. This is a rather
 conservative approach. If we had to consider that half of the DM is made
 of e.g. $S_1$ and the other half of $S_2$, some extra models with
 $m_{S_1}< m_{S_3}$ could be considered as viable.

Also notice that an important change in the
number of relativistic degrees of freedom $g_*$ (from $\sim$ 100 to
$\sim$ 10, see e.g. Ref.~\cite{Kolb}) occurs in the early universe
around\footnote{A $T_{QCD}\sim 150$ MeV has been taken into account
  in {\tt MicrOMEGAS } code} $T_{QCD}\sim 150$ MeV due to the QCD
phase transition. This implies larger values of the annihilation
cross-section for $m_{DM}\sim T_{QCD}\,x_f$. 

The relative absence of points in the $m_{DM}\sim m_{h}/2$ is
due to resonant DM annihilation through Higgs exchange. This feature
is also well visible in the $\Lambda_{DM}-m_{DM}$ and $\sigma_{DM\,
  p}-m_{DM}$ plots.  Another dip appears in all cases at $m_{DM}\sim
m_W$ when the annihilation into gauge bosons through Higgs exchange
become kinematically available. The other dips specific to the form of
each operators are discussed in the following.

Considering now the $\Lambda_{DM}-m_{DM}$, we see that above the Higgs
resonance the value of the $\Lambda_{DM}$ does not play an important
role given that any value above $ 100$ GeV is allowed by all the
constraints. Below the Higgs resonance, $\Lambda_{DM}$ should be
typically smaller than $\cO (100)$ GeV to account for the DM
relic abundance together with small contributions from the Higgs
portal to evade a large correction to the Higgs width. Let us
also remind that flavour observables, mainly meson decays into DM
particles, impose that $\Lambda_{DM}$ is larger than hundreds of GeV
for $m_{DM}$ in the GeV range~\cite{Kamenik:2011vy}. More precisely,
this constraint rules out $m_{DM}<m_{D}/2$ and $m_{DM}<m_{B}/2$ for
the operators $ {\cal O}_4$ and $ {\cal O}_{1,5}$, respectively and
the excluded regions are represented as (grey) shaded areas in
Figs.~\ref{fig:O123} and~\ref{fig:O45-s-m}. Moreover, the meson
oscillation tends to constrain $\Delta C^{ij}<0.1$, see
Sec.~\ref{sec:constraint}. This threatens the DM models of $\cO_5$ with
the lowest values of $\Lambda_{DM}$. The few points that are
eventually excluded by this constraint are represented in light blue
colors in the plots of the right column of Fig.~\ref{fig:O45-s-m}.

The candidates that pass direct and indirect detection searches
constraints (dark blue points) appear in two distinct mass ranges: below
10 GeV, where the direct detection bounds are weak, and above
$m_h/2$ as in the case of Higgs portal models~\cite{Djouadi:2011aa}:
\begin{itemize}
\item[-] For masses below 10 GeV, the scale of new physics is below
  $\sim 200$ GeV and the processes driving the relic abundance are 
  typically related to the dimension-6 operators $\cO_{1,..,5}$.

\item[-] Around $60$ GeV, near the Higgs resonance, the
  scale of new physics is typically above 1 TeV and the coupling to
  the Higgs particle is small ($\lambda<0.01$).

 \item[-] Above 60 GeV, the scale of new physics is of several
   hundreds of GeV and the coupling to the Higgs particle is
   $\cO(0.1-\pi)$.
\end{itemize}

\boldmath
\subsubsection{Operators $\cO_{1,2,3}$}
\label{sec:O123}
\unboldmath

In Fig.~\ref{fig:O123}, we present scatter plots associated to the
operator $\cO_1$.  We have checked that equivalent results for
$\cO_{2,3}$ can be obtained (up to some $\cO(1)$ factor in the
$\Lambda_{DM}$).  Given the p-wave suppression of the annihilation
cross-section, see Eq.~(\ref{eq:sv}), it is the direct detection
searches sector that provides the strongest constraints when $\cO_1$
drives the DM relic abundance. In the low mass regime, they are
limited by PICASSO's energy threshold at $m_{DM}\sim 4$ GeV. Let us
emphasise though that for scalar DM interacting through vector
interactions with quarks, it has been shown~\cite{Goodman:2010ku} that
monojet searches at Tevatron already provide very strong constraints
on GeV range DM masses and rule out all the new potentially viable DM
candidates with $m_{DM}< 10$ GeV, as shown in the plot on the right
hand side of Fig.~\ref{fig:O123}. For masses just above Higgs
resonance, the presence of extra interactions through $\cO_1$ allows
for lower value of the $\sdmp$ than in Higgs portal models and viable
candidates that could be within the reach of future Xenon 1T
experiments and in some rare cases will be tested by LHC at $\sqrt{s}=14$
TeV.

\boldmath
\subsubsection{Operators $\cO_{4,5}$}
\label{sec:O45}
\unboldmath

In the case of the $\cO_{4,5}$ operators, the results are shown in
Fig.~\ref{fig:O45-s-m}. This time the annihilation cross-section is an
s-wave process so that the indirect detection searches can compete
with direct detection bounds for $m_{DM}< 10-20$ GeV.  In the
$m_{DM}-\sv$ plane, in addition to the Higgs and gauge bosons
resonances, some other dips appear at $m_{DM}\sim m_c, m_t$
in the ${\cal O}_4$ and $m_{DM}\sim m_b$ in the ${\cal O}_5$
case. They are due to the fact that the main DM annihilation channels
into $S_i^*S_i\to\bar c c, \bar t t$ and $\bar bb$, respectively, are
suppressed by factors of $(-m_q^2 +m_{DM}^2)^{1/2}$ in the low
velocity limit, see Eq.~(\ref{eq:sv}).  Away from those dips and
resonances, the points getting lower values of the annihilation
cross-section have their relic abundance mainly driven by
coannihilation processes.

As it can be seen in the bottom plots of Fig.~\ref{fig:O45-s-m},
monojets searches~\cite{Goodman:2010ku} are not very limiting in the
case of scalar type DM-quarks interactions. The low mass regime appears
to pass all constraints in contrast to the case of simple Higgs portal
scenarios or to the case of Higgs portal + $\cO_{1,2,3}$ considered in
the previous section.

The particularities of operators ${\cal O}_4$ and
${\cal O}_5$ are the following:
\vspace{0.5cm}
\paragraph{Operator ${\cal O}_4$: }
In the low mass range, viable DM candidates get an extra suppression
of their relic abundance due to $S_1 S_2^* \to u \bar c$
coannihilation process, driven by the interaction terms proportional
to $c^4_2$ and $c^4_4$ in Eq.~(\ref{c4}). For $m_{1,2}<m_c$, the coannihilation
drives the relic density, while for $m_{1,2}>m_c$ it is a combination
of coannihilations and $S_2 S_2^* \to c \bar c$ that is
relevant.  
Let us remind that in order to get a non-negligible contribution from
coannihilations between two particles, their relative mass difference
should be $\Delta m/m \sim 0.1$ in order to avoid exponential
Boltzmann suppression.  In the case of $\cO_4$ no particular tuning in
$m_{1,2}$ has to be invoked given that their degeneracy is granted
thanks to the small $u$ and $c$ quarks Yukawa couplings in
Eq.~(\ref{spectrum}).\\ Also notice that in the case of ${\cal O}_4$,
for $m_{DM}>m_{t}$ regime, one can obtain a very reduced value of
$\sdmp$ compared to minimal Higgs portal scenarios (see
e.g.~\cite{Djouadi:2011aa}). This typically happens in the case of
negligible coupling to the Higgs so that $S_iS_i^*\rightarrow WW, ZZ,
\bar t t, hh$ through Higgs exchange are suppressed while ${\cal O}_4$
drives the DM relic abundance through $S_iS_i^*\rightarrow \bar t t$
for $m_{DM}>m_{t}$. This can be understood qualitatively by comparing
the DM-proton scattering cross-sections for Higgs portal and $\cO_4$
models.  In the first case $\sdmp^h\propto
f_{p\,h}^2\lambda_i^2/(m_h^4m_{DM}^2)\mu^2$~\cite{Andreas:2008xy}
while in the latter case $\sdmp^{\cO_4}\propto f_{p\,
  4}^2c^2/(\Lambda_{DM}^4m_{DM}^2)\mu^2$ ~\cite{Batell:2011tc} and
$\mu$ is the DM-proton reduced mass. The nucleon form factors
corresponds to $f_{p\,h}\simeq 0.4$ and $f_{p\,4}\simeq 0.1$ in the
Higgs portal and in the $\cO_4$ case, respectively, using the {\tt
  MicrOMEGAs} default parameters~\cite{Belanger:2008sj}. The
differences between $f_{p\, h}$ and $f_{p\, 4}$ are principally due to
the fact that $\cO_4$ does not provide a coupling to the $s$ quark. For
large DM masses, one can also relates the values of $\sv \propto
g^2\lambda^2/m_{DM}^2$ and $\sv \propto c^2 m_t^2/\Lambda_{DM}^4$ for
a DM abundance driven by Higgs portal or $\cO_4$ processes,
respectively. Eventually we get for $m_{DM}\ll m_t$, a $\sdmp^{\cO_4}$
that is suppressed compared to $\sdmp^{h}$ by a factor $(f_{p\,
  4}/f_{p\,h})^2$ and also by a factor $\sim m_h^4/(m_t^2m_{DM}^2)$.
\vspace{.5cm}

\paragraph{Operator ${\cal O}_5$:}

Below 10 GeV, viable candidates have their relic density mainly driven
by coannihilations $S_1 S_3^* \to \bar b d$ and $S_2 S_3^* \to \bar b
s$.  All these coannihilation channels are described by the flavour
interactions proportional to $c^5_2$ and $c^5_4$ in Eq.~(\ref{c5}). The $S_{1,2} S_3^*$ coannihilation processes require some tuning of
the parameters. Eq.~(\ref{spectrum}) gives rise to $m_1,
m_2\sim |m_A|$ and $m_3\sim |m_B|$ so that, if
$a^{(\prime)},b^{(\prime)}=+1$, $m_3\simeq 2 m_{1,2}$. Allowing for
different signs and the range of parameters in Eq.~(\ref{eq:ranges}),
we can get relative mass differences $\sim 0.1$ necessary for
coannihilation.

\vspace{.5cm} From the above analysis appears that the coannihilation
channels $S_1 S_2^* \to u \bar c$ for $\cO_4$ and $S_1 S_3^* \to \bar
b d$ and $S_2 S_3^* \to \bar b s$ for $\cO_5$ enable viable DM
parameter space in the GeV range. This is due to the flavour
interactions  driven by $c^\alpha_2$ and $c^\alpha_4$ with $\alpha= 4,5$ and this feature is absent in
standard Higgs portal models. Notice that the GeV mass range will be tested in the next future by direct detection searches experiment, see e.g. Ref.~\cite{MirabolfathiTalk} for SuperCDMS. Moreover, in the higher mass regime
$m_{DM}>m_t$, $\cO_4$ can give rise to rather suppressed DM-nucleon
cross-sections that could even evade future Xenon 1T constraints.

\section{Conclusions}
\label{sec:Conc}
Embedding the DM problematic within MFV context can guarantee DM
stability~\cite{Batell:2011tc}. In addition, flavour DM-fermions
interactions, that naturally arise in this framework, give rise to a
richer DM phenomenology than in the minimal Higgs portal scenarios
(see e.g.~\cite{Djouadi:2011aa} for a recent analysis).  In this
paper, we focused on the MFV in the quark sector only with the Yukawas
of the quarks being the only sources of flavour and CP violation. For
definiteness, we have restricted our analysis to the case of scalar DM
fields, neutral under the SM gauge symmetry, but transforming as a
triplet under one of the $SU(3)$ composing the flavour symmetry group
$G_f=SU(3)_{Q_L}\times SU(3)_{u_R}\times SU(3)_{d_R}$.

In this scenario, DM interacts with the SM fields through Higgs portal
at renormalisable level and also to quarks through dimension-6
operators. We have considered three vector operators and two scalar
ones. We have performed a systematic analysis of the DM viable
parameter space for each of these operators. We have obtained
that DM candidates can pass direct and indirect DM detection searches
constraints both below 10 GeV and above $m_h/2$. 

Complementary constraints from colliders and flavour physics have been
taken into account. In general the non observation of meson decays
into invisible final states exclude $m_{DM}<m_{D}/2$ or
$m_{DM}<m_{B}/2$ mass ranges. In the case of dimension-6 operators
with DM-quarks vector like interactions, monojet searches at colliders
rule out all the new potentially viable DM candidates with $m_{DM}<
10$ GeV. For what concerns scalar type dimension-6 interactions,
monojet searches are less limiting and allow for a viable $m_{DM}< 10$
GeV parameter space that will probably evade future LHC monojet
searches.  In most cases, it also evades constraints from
flavour violating processes such as meson oscillations. In the large
mass regime, most of the viable parameter space will be within the
reach of future direct detection searches experiments such as Xenon
1T, apart from the case of DM scalar-type coupling to $u$-type quarks
that can evade the latter bound for $m_{DM}> m_t$.

To summarise, when the DM is embedded in the MFV framework, the DM
stability is granted for certain representations of the DM fields
under $G_f$. In this work, we systematically analysed the rich
interplay between flavour and DM physics when considering vector and scalar type DM-quark interactions invariant under the original $G_f$ symmetry.  Compared with minimal Higgs portal
models, DM candidates with $m_{DM}\sim $ GeV range are viable in these scenarios. Furthermore, DM particles with $m_{DM}>m_t$ could even be beyond the reach of future
experiments such as Xenon 1T.

\section*{Acknowledgements}
We thank Jean-Marie Fr\`ere, Belen Gavela, Michael Gustafsson, Alejandro Ibarra, Gino Isidori, Stefano Rigolin, Emmanuel Stamou and Michel Tytgat for useful discussions. LLH is supported through an ``FWO-Vlaanderen'' post-doctoral fellowship project number 1271513. LLH also recognises partial support from the Belgian Federal Science Policy Office through the Interuniversity Attraction Pole P7/37 and from the Strategic Research Program ``High-Energy Physics'' of the Vrije Universiteit Brussel. LM acknowledges partial support by European Union FP7 ITN INVISIBLES (Marie Curie Actions, PITN-GA-2011-289442) and the Juan de la Cierva programme (JCI-2011-09244).

\providecommand{\href}[2]{#2}\begingroup\raggedright\endgroup


\begin{thebibliography}{10}

\bibitem{:2012gk}
{\bf ATLAS Collaboration} Collaboration, G.~Aad {\em et.~al.},  Phys.Lett. {\bf
  B716} (2012) 1--29, [\href{http://xxx.lanl.gov/abs/1207.7214}{{\tt
  arXiv:1207.7214}}].

\bibitem{:2012gu}
{\bf CMS Collaboration} Collaboration, S.~Chatrchyan {\em et.~al.},  Phys.Lett.
  {\bf B716} (2012) 30--61, [\href{http://xxx.lanl.gov/abs/1207.7235}{{\tt
  arXiv:1207.7235}}].

\bibitem{Patt:2006fw}
B.~Patt and F.~Wilczek,  \href{http://xxx.lanl.gov/abs/hep-ph/0605188}{{\tt
  hep-ph/0605188}}.

\bibitem{Kim:2006af}
Y.~G. Kim and K.~Y. Lee,  Phys.Rev. {\bf D75} (2007) 115012,
  [\href{http://xxx.lanl.gov/abs/hep-ph/0611069}{{\tt hep-ph/0611069}}].

\bibitem{MarchRussell:2008yu}
J.~March-Russell, S.~M. West, D.~Cumberbatch, and D.~Hooper,  JHEP {\bf 0807}
  (2008) 058, [\href{http://xxx.lanl.gov/abs/0801.3440}{{\tt
  arXiv:0801.3440}}].

\bibitem{Kim:2008pp}
Y.~G. Kim, K.~Y. Lee, and S.~Shin,  JHEP {\bf 0805} (2008) 100,
  [\href{http://xxx.lanl.gov/abs/0803.2932}{{\tt arXiv:0803.2932}}].

\bibitem{Ahlers:2008qc}
M.~Ahlers, J.~Jaeckel, J.~Redondo, and A.~Ringwald,  Phys.Rev. {\bf D78} (2008)
  075005, [\href{http://xxx.lanl.gov/abs/0807.4143}{{\tt arXiv:0807.4143}}].

\bibitem{Feng:2008mu}
J.~L. Feng, H.~Tu, and H.-B. Yu,  JCAP {\bf 0810} (2008) 043,
  [\href{http://xxx.lanl.gov/abs/0808.2318}{{\tt arXiv:0808.2318}}].

\bibitem{Andreas:2008xy}
S.~Andreas, T.~Hambye, and M.~H. Tytgat,  JCAP {\bf 0810} (2008) 034,
  [\href{http://xxx.lanl.gov/abs/0808.0255}{{\tt arXiv:0808.0255}}].

\bibitem{Barger:2008jx}
V.~Barger, P.~Langacker, M.~McCaskey, M.~Ramsey-Musolf, and G.~Shaughnessy,
  Phys.Rev. {\bf D79} (2009) 015018,
  [\href{http://xxx.lanl.gov/abs/0811.0393}{{\tt arXiv:0811.0393}}].

\bibitem{Kadastik:2009ca}
M.~Kadastik, K.~Kannike, A.~Racioppi, and M.~Raidal,  Phys.Rev.Lett. {\bf 104}
  (2010) 201301, [\href{http://xxx.lanl.gov/abs/0912.2729}{{\tt
  arXiv:0912.2729}}].

\bibitem{Kanemura:2010sh}
S.~Kanemura, S.~Matsumoto, T.~Nabeshima, and N.~Okada,  Phys.Rev. {\bf D82}
  (2010) 055026, [\href{http://xxx.lanl.gov/abs/1005.5651}{{\tt
  arXiv:1005.5651}}].

\bibitem{Piazza:2010ye}
F.~Piazza and M.~Pospelov,  Phys.Rev. {\bf D82} (2010) 043533,
  [\href{http://xxx.lanl.gov/abs/1003.2313}{{\tt arXiv:1003.2313}}].

\bibitem{Arina:2010an}
C.~Arina, F.-X. Josse-Michaux, and N.~Sahu,  Phys.Rev. {\bf D82} (2010) 015005,
  [\href{http://xxx.lanl.gov/abs/1004.3953}{{\tt arXiv:1004.3953}}].

\bibitem{Low:2011kp}
I.~Low, P.~Schwaller, G.~Shaughnessy, and C.~E. Wagner,  Phys.Rev. {\bf D85}
  (2012) 015009, [\href{http://xxx.lanl.gov/abs/1110.4405}{{\tt
  arXiv:1110.4405}}].

\bibitem{Djouadi:2011aa}
A.~Djouadi, O.~Lebedev, Y.~Mambrini, and J.~Quevillon,  Phys.Lett. {\bf B709}
  (2012) 65--69, [\href{http://xxx.lanl.gov/abs/1112.3299}{{\tt
  arXiv:1112.3299}}].

\bibitem{Englert:2011yb}
C.~Englert, T.~Plehn, D.~Zerwas, and P.~M. Zerwas,  Phys.Lett. {\bf B703}
  (2011) 298--305, [\href{http://xxx.lanl.gov/abs/1106.3097}{{\tt
  arXiv:1106.3097}}].

\bibitem{Kamenik:2012hn}
J.~F. Kamenik and C.~Smith,  Phys.Rev. {\bf D85} (2012) 093017,
  [\href{http://xxx.lanl.gov/abs/1201.4814}{{\tt arXiv:1201.4814}}].

\bibitem{Gonderinger:2012rd}
M.~Gonderinger, H.~Lim, and M.~J. Ramsey-Musolf,  Phys.Rev. {\bf D86} (2012)
  043511, [\href{http://xxx.lanl.gov/abs/1202.1316}{{\tt arXiv:1202.1316}}].

\bibitem{Lebedev:2012zw}
O.~Lebedev,  Eur.Phys.J. {\bf C72} (2012) 2058,
  [\href{http://xxx.lanl.gov/abs/1203.0156}{{\tt arXiv:1203.0156}}].

\bibitem{Farina:2011bh}
M.~Farina, M.~Kadastik, D.~Pappadopulo, J.~Pata, M.~Raidal, {\em et.~al.},
  Nucl.Phys. {\bf B853} (2011) 607--624,
  [\href{http://xxx.lanl.gov/abs/1104.3572}{{\tt arXiv:1104.3572}}].

\bibitem{Raidal:2011xk}
M.~Raidal and A.~Strumia,  Phys.Rev. {\bf D84} (2011) 077701,
  [\href{http://xxx.lanl.gov/abs/1108.4903}{{\tt arXiv:1108.4903}}].

\bibitem{Baek:2011aa}
S.~Baek, P.~Ko, and W.-I. Park,  JHEP {\bf 1202} (2012) 047,
  [\href{http://xxx.lanl.gov/abs/1112.1847}{{\tt arXiv:1112.1847}}].

\bibitem{Djouadi:2012zc}
A.~Djouadi, A.~Falkowski, Y.~Mambrini, and J.~Quevillon,
  \href{http://xxx.lanl.gov/abs/1205.3169}{{\tt arXiv:1205.3169}}.

\bibitem{Andreas:2010dz}
S.~Andreas, C.~Arina, T.~Hambye, F.-S. Ling, and M.~H. Tytgat,  Phys.Rev. {\bf
  D82} (2010) 043522, [\href{http://xxx.lanl.gov/abs/1003.2595}{{\tt
  arXiv:1003.2595}}].

\bibitem{Gustafsson:2012aj}
M.~Gustafsson, S.~Rydbeck, L.~Lopez-Honorez, and E.~Lundstrom,  Phys.Rev. {\bf
  D86} (2012) 075019, [\href{http://xxx.lanl.gov/abs/1206.6316}{{\tt
  arXiv:1206.6316}}].

\bibitem{Mambrini:2012ue}
Y.~Mambrini, M.~H. Tytgat, G.~Zaharijas, and B.~Zaldivar,  JCAP {\bf 1211}
  (2012) 038, [\href{http://xxx.lanl.gov/abs/1206.2352}{{\tt
  arXiv:1206.2352}}].

\bibitem{Wang:2012ts}
L.~Wang and X.-F. Han,  Phys.Rev. {\bf D87} (2013) 015015,
  [\href{http://xxx.lanl.gov/abs/1209.0376}{{\tt arXiv:1209.0376}}].

\bibitem{Hambye:2010zb}
T.~Hambye,  PoS {\bf IDM2010} (2011) 098,
  [\href{http://xxx.lanl.gov/abs/1012.4587}{{\tt arXiv:1012.4587}}].

\bibitem{Ma:2008ym}
E.~Ma,  Phys.Lett. {\bf B671} (2009) 366--368,
  [\href{http://xxx.lanl.gov/abs/0808.1729}{{\tt arXiv:0808.1729}}].

\bibitem{Hirsch:2010ru}
M.~Hirsch, S.~Morisi, E.~Peinado, and J.~Valle,  Phys.Rev. {\bf D82} (2010)
  116003, [\href{http://xxx.lanl.gov/abs/1007.0871}{{\tt arXiv:1007.0871}}].

\bibitem{Haba:2010ag}
N.~Haba, Y.~Kajiyama, S.~Matsumoto, H.~Okada, and K.~Yoshioka,  Phys.Lett. {\bf
  B695} (2011) 476--481, [\href{http://xxx.lanl.gov/abs/1008.4777}{{\tt
  arXiv:1008.4777}}].

\bibitem{Meloni:2010sk}
D.~Meloni, S.~Morisi, and E.~Peinado,  Phys.Lett. {\bf B697} (2011) 339--342,
  [\href{http://xxx.lanl.gov/abs/1011.1371}{{\tt arXiv:1011.1371}}].

\bibitem{Kajiyama:2010sb}
Y.~Kajiyama and H.~Okada,  Nucl.Phys. {\bf B848} (2011) 303--313,
  [\href{http://xxx.lanl.gov/abs/1011.5753}{{\tt arXiv:1011.5753}}].

\bibitem{Boucenna:2011tj}
M.~Boucenna, M.~Hirsch, S.~Morisi, E.~Peinado, M.~Taoso, {\em et.~al.},  JHEP
  {\bf 1105} (2011) 037, [\href{http://xxx.lanl.gov/abs/1101.2874}{{\tt
  arXiv:1101.2874}}].

\bibitem{Adulpravitchai:2011ei}
A.~Adulpravitchai, B.~Batell, and J.~Pradler,  Phys.Lett. {\bf B700} (2011)
  207--216, [\href{http://xxx.lanl.gov/abs/1103.3053}{{\tt arXiv:1103.3053}}].

\bibitem{Meloni:2011cc}
D.~Meloni, S.~Morisi, and E.~Peinado,  Phys.Lett. {\bf B703} (2011) 281--287,
  [\href{http://xxx.lanl.gov/abs/1104.0178}{{\tt arXiv:1104.0178}}].

\bibitem{deAdelhartToorop:2011ad}
R.~de~Adelhart~Toorop, F.~Bazzocchi, and S.~Morisi,  Nucl. Phys. {\bf B856}
  (2012) 670--681, [\href{http://xxx.lanl.gov/abs/1104.5676}{{\tt
  arXiv:1104.5676}}].

\bibitem{Lavoura:2011ry}
L.~Lavoura,  J.Phys. {\bf G39} (2012) 025202,
  [\href{http://xxx.lanl.gov/abs/1109.6854}{{\tt arXiv:1109.6854}}].

\bibitem{Hashimoto:2011tn}
K.~Hashimoto and H.~Okada,  \href{http://xxx.lanl.gov/abs/1110.3640}{{\tt
  arXiv:1110.3640}}.

\bibitem{Boucenna:2012qb}
M.~Boucenna, S.~Morisi, E.~Peinado, Y.~Shimizu, and J.~Valle,  Phys.Rev. {\bf
  D86} (2012) 073008, [\href{http://xxx.lanl.gov/abs/1204.4733}{{\tt
  arXiv:1204.4733}}].

\bibitem{Lavoura:2012cv}
L.~Lavoura, S.~Morisi, and J.~Valle,  JHEP {\bf 1302} (2013) 118,
  [\href{http://xxx.lanl.gov/abs/1205.3442}{{\tt arXiv:1205.3442}}].

\bibitem{BenTov:2012tg}
Y.~BenTov, X.-G. He, and A.~Zee,  \href{http://xxx.lanl.gov/abs/1208.1062}{{\tt
  arXiv:1208.1062}}.

\bibitem{Bhattacharya:2013mpa}
S.~Bhattacharya, E.~Ma, A.~Natale, and A.~Rashed,
  \href{http://xxx.lanl.gov/abs/1302.6266}{{\tt arXiv:1302.6266}}.

\bibitem{Hamaguchi:1998wm}
K.~Hamaguchi, Y.~Nomura, and T.~Yanagida,  Phys.Rev. {\bf D58} (1998) 103503,
  [\href{http://xxx.lanl.gov/abs/hep-ph/9805346}{{\tt hep-ph/9805346}}].

\bibitem{Hamaguchi:1998nj}
K.~Hamaguchi, Y.~Nomura, and T.~Yanagida,  Phys.Rev. {\bf D59} (1999) 063507,
  [\href{http://xxx.lanl.gov/abs/hep-ph/9809426}{{\tt hep-ph/9809426}}].

\bibitem{Batell:2010bp}
B.~Batell,  Phys.Rev. {\bf D83} (2011) 035006,
  [\href{http://xxx.lanl.gov/abs/1007.0045}{{\tt arXiv:1007.0045}}].

\bibitem{Blennow:2010qp}
M.~Blennow, B.~Dasgupta, E.~Fernandez-Mart{\'\i ne}z, and N.~Rius,  JHEP {\bf
  1103} (2011) 014, [\href{http://xxx.lanl.gov/abs/1009.3159}{{\tt
  arXiv:1009.3159}}].

\bibitem{Batell:2011tc}
B.~Batell, J.~Pradler, and M.~Spannowsky,  JHEP {\bf 1108} (2011) 038,
  [\href{http://xxx.lanl.gov/abs/1105.1781}{{\tt arXiv:1105.1781}}].

\bibitem{Chivukula:1987py}
R.~S. Chivukula and H.~Georgi,  Phys.Lett. {\bf B188} (1987) 99.

\bibitem{Hall:1990ac}
L.~J. Hall and L.~Randall,  Phys. Rev. Lett. {\bf 65} (1990) 2939--2942.

\bibitem{D'Ambrosio:2002ex}
G.~D'Ambrosio, G.~Giudice, G.~Isidori, and A.~Strumia,  Nucl.Phys. {\bf B645}
  (2002) 155--187, [\href{http://xxx.lanl.gov/abs/hep-ph/0207036}{{\tt
  hep-ph/0207036}}].

\bibitem{Cirigliano:2005ck}
V.~Cirigliano, B.~Grinstein, G.~Isidori, and M.~B. Wise,  Nucl.Phys. {\bf B728}
  (2005) 121--134, [\href{http://xxx.lanl.gov/abs/hep-ph/0507001}{{\tt
  hep-ph/0507001}}].

\bibitem{Davidson:2006bd}
S.~Davidson and F.~Palorini,  Phys.Lett. {\bf B642} (2006) 72--80,
  [\href{http://xxx.lanl.gov/abs/hep-ph/0607329}{{\tt hep-ph/0607329}}].

\bibitem{Gavela:2009cd}
M.~Gavela, T.~Hambye, D.~Hernandez, and P.~Hernandez,  JHEP {\bf 0909} (2009)
  038, [\href{http://xxx.lanl.gov/abs/0906.1461}{{\tt arXiv:0906.1461}}].

\bibitem{Feldmann:2009dc}
T.~Feldmann, M.~Jung, and T.~Mannel,  Phys.Rev. {\bf D80} (2009) 033003,
  [\href{http://xxx.lanl.gov/abs/0906.1523}{{\tt arXiv:0906.1523}}].

\bibitem{Alonso:2011yg}
R.~Alonso, M.~Gavela, L.~Merlo, and S.~Rigolin,  JHEP {\bf 1107} (2011) 012,
  [\href{http://xxx.lanl.gov/abs/1103.2915}{{\tt arXiv:1103.2915}}].

\bibitem{Alonso:2011jd}
R.~Alonso, G.~Isidori, L.~Merlo, L.~A. Munoz, and E.~Nardi,  JHEP {\bf 1106}
  (2011) 037, [\href{http://xxx.lanl.gov/abs/1103.5461}{{\tt
  arXiv:1103.5461}}].

\bibitem{Alonso:2012fy}
R.~Alonso, M.~Gavela, D.~Hernandez, and L.~Merlo,  Phys.Lett. {\bf B715} (2012)
  194--198, [\href{http://xxx.lanl.gov/abs/1206.3167}{{\tt arXiv:1206.3167}}].

\bibitem{Fitzpatrick:2007sa}
A.~L. Fitzpatrick, G.~Perez, and L.~Randall,  Phys.Rev.Lett. {\bf 100} (2008)
  171604, [\href{http://xxx.lanl.gov/abs/0710.1869}{{\tt arXiv:0710.1869}}].

\bibitem{Kagan:2009bn}
A.~L. Kagan, G.~Perez, T.~Volansky, and J.~Zupan,  Phys. Rev. {\bf D80} (2009)
  076002, [\href{http://xxx.lanl.gov/abs/0903.1794}{{\tt arXiv:0903.1794}}].

\bibitem{Lalak:2010bk}
Z.~Lalak, S.~Pokorski, and G.~G. Ross,  JHEP {\bf 1008} (2010) 129,
  [\href{http://xxx.lanl.gov/abs/1006.2375}{{\tt arXiv:1006.2375}}].

\bibitem{Grinstein:2010ve}
B.~Grinstein, M.~Redi, and G.~Villadoro,  JHEP {\bf 1011} (2010) 067,
  [\href{http://xxx.lanl.gov/abs/1009.2049}{{\tt arXiv:1009.2049}}].

\bibitem{Buras:2011wi}
A.~J. Buras, M.~V. Carlucci, L.~Merlo, and E.~Stamou,  JHEP {\bf 1203} (2012)
  088, [\href{http://xxx.lanl.gov/abs/1112.4477}{{\tt arXiv:1112.4477}}].

\bibitem{Alonso:2012jc}
R.~Alonso, M.~Gavela, L.~Merlo, S.~Rigolin, and J.~Yepes,  JHEP {\bf 1206}
  (2012) 076, [\href{http://xxx.lanl.gov/abs/1201.1511}{{\tt
  arXiv:1201.1511}}].

\bibitem{Alonso:2012px}
R.~Alonso, M.~Gavela, L.~Merlo, S.~Rigolin, and J.~Yepes,
  \href{http://xxx.lanl.gov/abs/1212.3305}{{\tt arXiv:1212.3305}}.

\bibitem{Alonso:2012pz}
R.~Alonso, M.~Gavela, L.~Merlo, S.~Rigolin, and J.~Yepes,
  \href{http://xxx.lanl.gov/abs/1212.3307}{{\tt arXiv:1212.3307}}.

\bibitem{Arnold:2009ay}
J.~M. Arnold, M.~Pospelov, M.~Trott, and M.~B. Wise,  JHEP {\bf 1001} (2010)
  073, [\href{http://xxx.lanl.gov/abs/0911.2225}{{\tt arXiv:0911.2225}}].

\bibitem{Agrawal:2011ze}
P.~Agrawal, S.~Blanchet, Z.~Chacko, and C.~Kilic,  Phys.Rev. {\bf D86} (2012)
  055002, [\href{http://xxx.lanl.gov/abs/1109.3516}{{\tt arXiv:1109.3516}}].

\bibitem{Kamenik:2011nb}
J.~F. Kamenik and J.~Zupan,  Phys.Rev. {\bf D84} (2011) 111502,
  [\href{http://xxx.lanl.gov/abs/1107.0623}{{\tt arXiv:1107.0623}}].

\bibitem{Isidori:2010kg}
G.~Isidori, Y.~Nir, and G.~Perez,  Ann. Rev. Nucl. Part. Sci. {\bf 60} (2010)
  355, [\href{http://xxx.lanl.gov/abs/1002.0900}{{\tt arXiv:1002.0900}}].

\bibitem{DelNobile:2011uf}
E.~Del~Nobile and F.~Sannino,  Int.J.Mod.Phys. {\bf A27} (2012) 1250065,
  [\href{http://xxx.lanl.gov/abs/1102.3116}{{\tt arXiv:1102.3116}}].

\bibitem{Kamenik:2011vy}
J.~F. Kamenik and C.~Smith,  JHEP {\bf 1203} (2012) 090,
  [\href{http://xxx.lanl.gov/abs/1111.6402}{{\tt arXiv:1111.6402}}].

\bibitem{Griest:1990kh}
K.~Griest and D.~Seckel,  Phys.Rev. {\bf D43} (1991) 3191--3203.

\bibitem{Hinshaw:2012fq}
G.~Hinshaw, D.~Larson, E.~Komatsu, D.~Spergel, C.~Bennett, {\em et.~al.},
  \href{http://xxx.lanl.gov/abs/1212.5226}{{\tt arXiv:1212.5226}}.

\bibitem{Belanger:2008sj}
G.~Belanger, F.~Boudjema, A.~Pukhov, and A.~Semenov,  Comput.Phys.Commun. {\bf
  180} (2009) 747--767, [\href{http://xxx.lanl.gov/abs/0803.2360}{{\tt
  arXiv:0803.2360}}].

\bibitem{Belanger:2010gh}
G.~Belanger, F.~Boudjema, P.~Brun, A.~Pukhov, S.~Rosier-Lees, {\em et.~al.},
  Comput.Phys.Commun. {\bf 182} (2011) 842--856,
  [\href{http://xxx.lanl.gov/abs/1004.1092}{{\tt arXiv:1004.1092}}].

\bibitem{Archambault:2012pm}
{\bf PICASSO Collaboration} Collaboration, S.~Archambault {\em et.~al.},
  Phys.Lett. {\bf B711} (2012) 153--161,
  [\href{http://xxx.lanl.gov/abs/1202.1240}{{\tt arXiv:1202.1240}}].

\bibitem{Aprile:2012nq}
{\bf XENON100 Collaboration} Collaboration, E.~Aprile {\em et.~al.},
  \href{http://xxx.lanl.gov/abs/1207.5988}{{\tt arXiv:1207.5988}}.

\bibitem{Ackermann:2011wa}
{\bf Fermi-LAT collaboration} Collaboration, M.~Ackermann {\em et.~al.},
  Phys.Rev.Lett. {\bf 107} (2011) 241302,
  [\href{http://xxx.lanl.gov/abs/1108.3546}{{\tt arXiv:1108.3546}}].

\bibitem{Hou:2012xq}
Z.~Hou, C.~Reichardt, K.~Story, B.~Follin, R.~Keisler, {\em et.~al.},
  \href{http://xxx.lanl.gov/abs/1212.6267}{{\tt arXiv:1212.6267}}.

\bibitem{Planck:2006aa}
{\bf Planck Collaboration} Collaboration,
  \href{http://xxx.lanl.gov/abs/astro-ph/0604069}{{\tt astro-ph/0604069}}.

\bibitem{Padmanabhan:2005es}
N.~Padmanabhan and D.~P. Finkbeiner,  Phys.Rev. {\bf D72} (2005) 023508,
  [\href{http://xxx.lanl.gov/abs/astro-ph/0503486}{{\tt astro-ph/0503486}}].

\bibitem{Natarajan:2009bm}
A.~Natarajan and D.~J. Schwarz,  Phys.Rev. {\bf D80} (2009) 043529,
  [\href{http://xxx.lanl.gov/abs/0903.4485}{{\tt arXiv:0903.4485}}].

\bibitem{Cirelli:2009bb}
M.~Cirelli, F.~Iocco, and P.~Panci,  JCAP {\bf 0910} (2009) 009,
  [\href{http://xxx.lanl.gov/abs/0907.0719}{{\tt arXiv:0907.0719}}].

\bibitem{Finkbeiner:2011dx}
D.~P. Finkbeiner, S.~Galli, T.~Lin, and T.~R. Slatyer,  Phys.Rev. {\bf D85}
  (2012) 043522, [\href{http://xxx.lanl.gov/abs/1109.6322}{{\tt
  arXiv:1109.6322}}].

\bibitem{Giesen:2012rp}
G.~Giesen, J.~Lesgourgues, B.~Audren, and Y.~Ali-Haimoud,
  \href{http://xxx.lanl.gov/abs/1209.0247}{{\tt arXiv:1209.0247}}.

\bibitem{Cline:2013fm}
J.~M. Cline and P.~Scott,  \href{http://xxx.lanl.gov/abs/1301.5908}{{\tt
  arXiv:1301.5908}}.

\bibitem{Giardino:2012dp}
P.~P. Giardino, K.~Kannike, M.~Raidal, and A.~Strumia,  Phys.Lett. {\bf B718}
  (2012) 469--474, [\href{http://xxx.lanl.gov/abs/1207.1347}{{\tt
  arXiv:1207.1347}}].

\bibitem{Chatrchyan:2012me}
{\bf CMS Collaboration} Collaboration, S.~Chatrchyan {\em et.~al.},  JHEP {\bf
  1209} (2012) 094, [\href{http://xxx.lanl.gov/abs/1206.5663}{{\tt
  arXiv:1206.5663}}].

\bibitem{ATLAS:2012ky}
{\bf ATLAS Collaboration} Collaboration, G.~Aad {\em et.~al.},
  \href{http://xxx.lanl.gov/abs/1210.4491}{{\tt arXiv:1210.4491}}.

\bibitem{Beltran:2008xg}
M.~Beltran, D.~Hooper, E.~W. Kolb, and Z.~C. Krusberg,  Phys.Rev. {\bf D80}
  (2009) 043509, [\href{http://xxx.lanl.gov/abs/0808.3384}{{\tt
  arXiv:0808.3384}}].

\bibitem{Beltran:2010ww}
M.~Beltran, D.~Hooper, E.~W. Kolb, Z.~A. Krusberg, and T.~M. Tait,  JHEP {\bf
  1009} (2010) 037, [\href{http://xxx.lanl.gov/abs/1002.4137}{{\tt
  arXiv:1002.4137}}].

\bibitem{Goodman:2010ku}
J.~Goodman, M.~Ibe, A.~Rajaraman, W.~Shepherd, T.~M. Tait, {\em et.~al.},
  Phys.Rev. {\bf D82} (2010) 116010,
  [\href{http://xxx.lanl.gov/abs/1008.1783}{{\tt arXiv:1008.1783}}].

\bibitem{Isidori:2013ez}
G.~Isidori,  \href{http://xxx.lanl.gov/abs/1302.0661}{{\tt arXiv:1302.0661}}.

\bibitem{Semenov:2008jy}
A.~Semenov,  Comput.Phys.Commun. {\bf 180} (2009) 431--454,
  [\href{http://xxx.lanl.gov/abs/0805.0555}{{\tt arXiv:0805.0555}}].

\bibitem{Xenon1T}
M.~Schumann,  2012.
\newblock Slides of a talk given at the SPS Meeting 2012, ETH Zurich,
  21.06.2012.

\bibitem{Kolb}
E.~Kolb and M.~Turner, {\em The Early Universe (Frontiers of Physics)}.
\newblock Addison-Wesley Publishing Company, 1990.

\bibitem{MirabolfathiTalk}
N.~Mirabolfathi.
\newblock Talk given at the Rencontres de Moriond 2013.

\end{thebibliography}
\end{document}